\documentclass[sigconf]{acmart}

 \renewcommand\footnotetextcopyrightpermission[1]{}
\AtBeginDocument{%
  \providecommand\BibTeX{{%
    \normalfont B\kern-0.5em{\scshape i\kern-0.25em b}\kern-0.8em\TeX}}}

\graphicspath{{Figures/}{../Figures/}}
\usepackage{subcaption}
\usepackage{mathrsfs}

\newif\ifshowcomments
\showcommentstrue

\ifshowcomments
\newcommand{\lakshnote}[2]{\textcolor{blue}{\fbox{\bfseries\sffamily\scriptsize#1}}
  \textcolor{blue}{{$/*$\textsf{\emph{#2}}$*/$}}}
\newcommand{\mohammadnote}[2]{\textcolor{red}{\fbox{\bfseries\sffamily\scriptsize#1}}
  \textcolor{red}{{$/*$\textsf{\emph{#2}}$*/$}}}

\else
\newcommand{\mohammadnote}[2]{}
\newcommand{\lakshnote}[2]{}
\fi

\usepackage{textcomp}
\usepackage{xcolor}
\usepackage{xspace}
\usepackage{enumitem}
\setitemize{noitemsep,topsep=0pt,parsep=0pt,partopsep=0pt}
\usepackage{tabularx}
\usepackage{multirow}
\usepackage{multicol}
\newcolumntype{L}{>{\raggedright\arraybackslash}X}
\newcommand{\AlgoName}{RFTacho\xspace}
\newcommand{\fakeparagraph}[1]{\vspace{1mm}\noindent{\textbf{#1}}}

\begin{document}

\title{RFTacho: Non-intrusive RF monitoring of rotating machines}

\author{Mohammad Heggo, Laksh Bhatia and Julie A. McCann}
\affiliation{\institution{Imperial College London}
\country{United Kingdom}}
\email{{m.heggo,laksh.bhatia16,j.mccann}@imperial.ac.uk}

\begin{abstract}
Measuring rotation speed is essential to many engineering applications; it elicits faults undetectable by vibration monitoring alone and enhances the vibration signal analysis of rotating machines. Optical, magnetic or mechanical Tachometers are currently state-of-art. Their limitations are they require line-of-sight, direct access to the rotating object. This paper proposes \AlgoName, a rotation speed measurement \emph{system} that leverages novel hardware and signal processing algorithms to produce highly accurate readings conveniently. \AlgoName uses RF Orbital Angular Momentum (OAM) waves to measure rotation speed of multiple machines simultaneously with no requirements from the machine's properties. OAM antennas allow it to operate in high-scattering environments, commonly found in industries, as they are resilient to de-polarization compared to linearly polarized antennas.
\AlgoName achieves this by using two novel signal processing algorithms to extract the rotation speed of several rotating objects simultaneously amidst noise arising from high-scattering environments, non-line-of-sight scenarios and dynamic environmental conditions with a resolution of $1 rpm$. We test \AlgoName on several real-world machines like fans, motors, air conditioners. Results show that \AlgoName has avg. error of $<0.5\%$ compared to ground truth. We demonstrate \AlgoName's simultaneous multiple-object measurement capability that other tachometers do not have. Initial experiments show that \AlgoName can measure speeds as high as 7000 rpm (theoretically 60000 rpm) with high resiliency at different coverage distances and orientation angles, requiring only 150 mW transmit power while operating in the 5 GHz license-exempt band. 
\AlgoName is the first RF-based sensing system that combines OAM waves and novel processing approaches to measure the rotation speed of multiple machines simultaneously in a non-intrusive way.

\end{abstract}


\maketitle
\begin{figure}
\includegraphics[width=0.47\textwidth]{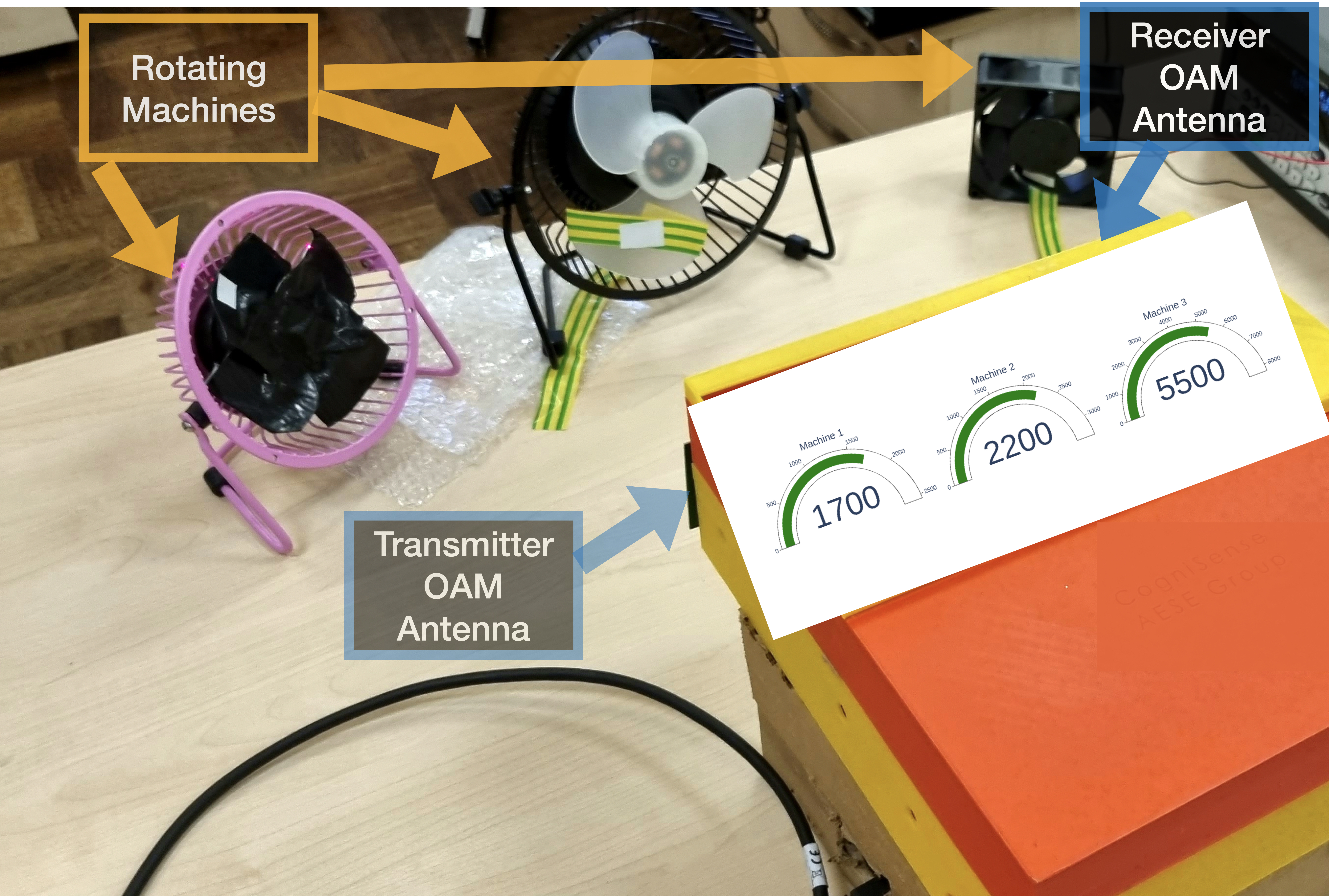}
\caption{\AlgoName antennas transmit OAM waves towards the machine rotor and receive its scattered wave. \AlgoName extracts the rotor speed by analysing the scattered wave using advanced signal processing}
\label{fig:workingprinciple}
\vspace{-5mm}
\end{figure}

\section{Introduction}

Condition monitoring of rotating machines has traditionally made use of vibration-based sensing, yet this is not ideal. The vibration signal from rotating machines exhibits a cyclo-stationary behaviour coupled to the rotation cycle of the machine \cite{he2016novel,leclere2016multi}, implying that rotation speed is required to enhance the vibration signal's extraction and analysis \cite{madhavan2014vibration,gubran2014shaft,liu2020review}.
In addition, speed information enables the detection of local incipient internal component faults, such as gear teeth wear or cracks, which are undetectable using vibration sensors alone\cite{cheng2020incipient}. As a result, the rotating machine's speed information is crucial to understanding nominal performance or to predict possible future failures \cite{castellani2020diagnosis,natili2020video}.

Tachometers are state-of-art for accurate rotation speed measurement. Mechanical tachometers measure rotor speed based on direct contact between the rotating shaft and the tachometer \cite{liptak2003instrument}. Optical tachometers measure speed by emitting a laser or infrared beam towards a rotating object that is fixed with a reflective tape \cite{liptak2003instrument}. The tape reflects the beam to the tachometer every rotation cycle, allowing rotation speed calculation. Magnetic (or hall-effect) tachometers measure the rotation speed of gears or slotted wheels \cite{liptak2003instrument} using changes to its analogue output signal voltage. This change arises from the change in the external magnetic field due to the discontinuous ferrite rotating object (i.e. the gear tooth). 

The technologies above have five constraints that make them non-ideal for cost-effective, practical deployment. The first is that mechanical and magnetic technologies require direct contact or a short separation distance (mm). Second, a line of sight path needs to exist between the tachometer and the rotating shaft.
Third, mounting or installing a sensor, reflective tape, or a tag on the machine body is impractical; since this maintenance activity increases machine stoppage hours.
Fourth, foreknowledge of the machine shape (e.g. number of rotating blades) is required in magnetic tachometers.
Finally, magnetic tachometers require rotating objects to be built with ferrite materials.

These constraints motivate the need for a low-cost, easy-to-deploy system that can accurately measure the rotation speed of an object without machine stoppage or foreknowledge about the machine. This paper proposes \AlgoName, a non-intrusive RF machine monitoring system that exploits the principle of rotational frequency Doppler shift to measure the rotation speed of multiple machines simultaneously. RF waves undergo rotational Doppler shift from the relative rotational motion between a spinning object and an RF transmitter that emits electromagnetic waves with a non-zero orbital angular momentum (OAM) wave\cite{zhao2016measurement}.

\fakeparagraph{Challenges:} Measuring rotation speed for multiple machines simultaneously is a non-trivial task and has several real-world challenges. \AlgoName tackles these challenges by leveraging novel hardware and lightweight signal processing algorithms. The three main challenges for \AlgoName and how they are addressed are: 

\fakeparagraph{1. High-Scattering Environments (Sec. \ref{antennacomparison}):} \AlgoName needs to operate in industrial scenarios, where radio waves undergo high scattering from heavy metallic machinery, that change the polarization of the transmitted electromagnetic fields\cite{cuinas2009depolarization}. In these environments, linearly polarized antennas (commonly used in communication systems) suffer from severe depolarization that makes it challenging to identify the signal from noise\cite{golmohamadi2016depolarization}. Depolarization further reduces the received signal power, leading to a significant decrease in the signal-to-noise ratio. For this reason, \AlgoName uses novel OAM antennas and waves. Presently, OAM waves have been only used for communication; \AlgoName is the first RF sensing system that uses OAM waves to measure rotation speed of machines. OAM waves are also immune against depolarization in high scattering environments, making them ideal for RF sensing applications.

\fakeparagraph{2. RF Speed Measurement (Sec.~\ref{speedextraction}):} 
\AlgoName's second challenge is estimating the main Doppler frequency shift (henceforth referred as $\mathscr{D}$-frequency) amidst noise and RF reflections from other machines and the environment. \AlgoName also needs to have a high-resolution of 1 rpm, commonly needed for industrial tasks. 
\AlgoName solves these challenges with a two-stage approach, a coarse speed measurement that estimates the speed only based on the $\mathscr{D}$-frequency and its harmonics and a fine speed measurement that uses information about signal amplitudes to achieve a higher resolution. The sensing resolution for \AlgoName is proportional to the sampling time as $1/60$ Hz. \AlgoName achieves the 1 rpm resolution by having a sampling time of $1$ second. The sampling time can be reduced to decrease \AlgoName's resolution and is an application-dependent parameter.  

\fakeparagraph{3. RF high background and narrowband noise interference (Sec.~\ref{multiplesubcarriers}):} The third challenge for \AlgoName is operation in dynamic environments where people and object movement is common. These environments lead to narrowband interference which affect the reflected OAM signals on a single subcarrier.
Industrial environments are also characterized by a high background noise that results from high temperatures and  switching modes of machines\cite{cheffena2016propagation}. \AlgoName tackles these challenges by leveraging multiple subcarriers distributed over a wider bandwidth. \AlgoName measures the speed of the machines over all subcarriers and reports the level of convergence (LoC). LoC is the number of subcarriers that report readings within $\pm 60$ rpm of the median measurements of all subcarriers. The number of subcarriers is a user-defined parameter and impacts the LoC. Increasing the number of subcarriers allows a high LoC but reduces the sensing range as the transmission power is evenly distributed among all subcarriers. \AlgoName's LoC is also helpful when choosing the appropriate threshold and transmit powers.

Fig.~\ref{fig:workingprinciple} shows \AlgoName's working principle. The hardware platform integrates the RF transmitter, receiver and processing system into a single monostatic-radar configuration. This configuration allows \AlgoName to be deployed with flexibility in industrial environments. \AlgoName's transceiver has been specifically designed to emit and receive OAM waves. It can measure the rotation speeds in non-line-of-sight scenarios, tilt between the sensing system and the test machine, and high-scattering environments. \AlgoName places no requirement on the type of machine, the number of blades, or rotor blades' composition, making it highly applicable to a significant range of industrial applications. We test \AlgoName in all the above scenarios and show how its performance is comparable to traditional tachometers but offers benefits like measuring the speed of multiple machines and non-line-of-sight sensing that tachometers do not have.

\fakeparagraph{Contributions.} \AlgoName is the first RF-based novel hardware-software sensing system that can measure the rotation speed of multiple machines simultaneously. \AlgoName uses OAM waves that allow it to operate in high-scattering environments and have lightweight signal processing algorithms, allowing for many deployment configurations.

\section{Background and Preliminaries}
\label{sec:background}

In this section, we will discuss the background of orbital angular momentum waves and rotation Doppler shift, the concepts that we use to sense the rotation speed of machines. We will also discuss the differences between linearly polarized antennas and OAM antennas and show why \AlgoName uses OAM antennas instead of the commonly found linear antennas.

\subsection{Orbital Angular Momentum Waves}

\begin{figure}
\begin{subfigure}{0.13\textwidth}
  \centering
    \includegraphics[height=3cm]{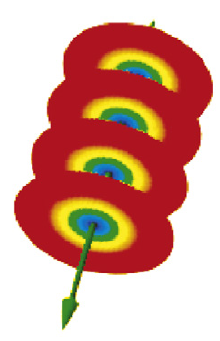}
    \caption{\label{fig:Planewave}}
\end{subfigure}%
\begin{subfigure}{0.13\textwidth}
  \centering
    \includegraphics[height=3cm]{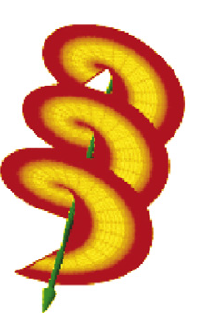}
    \caption{\label{fig:OAMl1}}
\end{subfigure}%
\begin{subfigure}{0.13\textwidth}
  \centering
    \includegraphics[height=3cm]{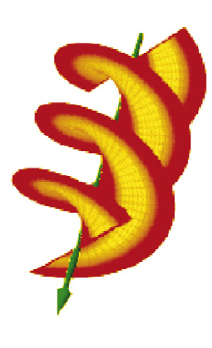}
    \caption{\label{fig:OAMl2}}
\end{subfigure}%
\caption{Wavefront of plane waves vs. OAM waves \cite{emile2017energy} (a) Plane wave (b) OAM wave with topological charge $l=1$ (c) OAM wave with topological charge $l=2$}
\label{OAMPlanewaves}
\end{figure}

Electromagnetic (EM) waves carry both linear and angular momentum \cite{chen2019orbital,zhang2020review}. The momentum of the EM field determines the radiation pressure that the wave creates upon incidence on an object. The angular momentum is divided into spin angular momentum (SAM) and orbital angular momentum (OAM). The SAM defines the polarization of the EM waves, as is the case in circularly polarized waves. The OAM identifies a wave's spatial rotation about a vortex \cite{chen2019orbital,zhang2020review}. The main difference between plane waves and OAM waves is that the latter experience a helical wavefront that is spatially connected, and the former has a wavefront of parallel separated rings in the space. The difference between the wavefront of the plane and OAM waves is shown in Fig. \ref{OAMPlanewaves}\cite{emile2017energy}.

OAM waves were first observed in 1992 by \citeauthor{allen1992orbital} in optical beams \cite{allen1992orbital}, where they were shown that Laguerre–Gaussian (LG) beams possess an OAM value of $l\hbar$ per photon. $\hbar$ is Plank's constant divided by $2\pi$, and $l$ represents the topological charge, which is the degree of the helical twist of the EM beam around the vortex. 
Since then, OAM has become an essential part of many applications such as optical tweezers, quantum entanglement, optical communication, non-linear optics, microscopy, and imaging \cite{shen2019optical,zhang2020review}. For example, in optical tweezers, the OAM of the optical beam is used to manipulate the angular position of particles. In optical communications, OAM optical beams with different topological charges carry parallel streams of information simultaneously, giving an alternative degree of multiplexing modulation. Moreover, OAM optical beams can detect the spinning motions based on the coupling between OAM and mechanical momentum. OAM rotation motion can be detected at nanoscales (molecules) or large scale (atmospheric turbulence or even rotating black holes).

In 2007, \citeauthor{thide2007utilization} generated OAM waves in radiofrequency (RF), which was then applied to wireless communication\cite{thide2007utilization}. Similar to the optical field, OAM waves have been used as an alternate multiplexing mode in wireless communications \cite{yan2014high,hui2015multiplexed}. Moreover, OAM waves have also been used in object identification \cite{uribe2013object}, and radar applications to detect object rotational motion \cite{lin2016improved}. However, these have been in lab scenarios and ignored the real-world problems of designing a practical sensing system.

\subsection{Rotation Doppler Shift}

Linear Doppler shift arises due to translational motion between an object and an EM wave source\cite{asakura1981dynamic}. However, rotational motion between an object and the wave source can induce a rotational Doppler shift, which was first discovered by \citeauthor{courtial1998rotational} in \cite{courtial1998rotational}. \citeauthor{courtial1998rotational} suggested that the LG optical beam that was previously proved to carry OAM by \citeauthor{allen1992orbital}\cite{allen1992orbital} exhibits a change in its frequency upon incidence on a rotating object. This change in the frequency was known as  \textit{rotational Doppler shift}. Since then Rotational Doppler shift of an optical OAM beam has been used to detect the rotation speed of objects on the macroscopic or atomic scale\cite{phillips2014rotational}.

The measurement of the rotational Doppler frequency in RF was first introduced in \cite{zhao2016measurement}, where Doppler frequency shift $\Delta f$ was evaluated as:
\begin{equation}
    \Delta f = {(\omega l)}/{(2\pi)}
    \label{equation1}
\end{equation}
where $l$ is the topological charge, and $\omega$ is the rotation speed of the spinning object. $l$ represents the phase front variation in a period of $2\pi$, and it has an integer value. Hence, in the case of plane waves (linearly or circularly polarized) where  $l = 0$, the rotational Doppler frequency shift is equal to zero \cite{zheng2018analysis}.
In OAM wave case ($l \neq 0$), we can calculate the speed of the rotating machine as:
\begin{equation}
    \omega_{rpm} = {(60 \times \Delta f)}/{l}
    \label{equation2}
\end{equation}

\subsection{Antenna Comparison}
\label{antennacomparison}

The antenna is a key component in \AlgoName's hardware as it is responsible for emitting and receiving the electromagnetic waves. However, linear antennas suffer from severe depolarization in industrial environments\cite{golmohamadi2016depolarization,cuinas2009depolarization}. Depolarization means that an obstacle alters the polarization of the incident electromagnetic wave by a certain percentage. This change in the scattered wave polarization leads to a significant loss of received power by linearly polarized receivers. In an industrial high scattering environment, the depolarization problem also leads to deep signal fading. This decrease in the received signal to noise ratio will lead to the wrong estimation of the machine rotation speed and require complex signal processing algorithms.

To overcome this challenge, \AlgoName uses OAM antennas in its transceiver. The OAM antennas can emit and receive electromagnetic waves that have a helical wavefront. 
This feature allows the OAM antennas to accommodate any changes in the electric field polarization direction in a high scattering industrial environment.

\begin{figure}
\begin{center}
\includegraphics[width=0.47\textwidth]{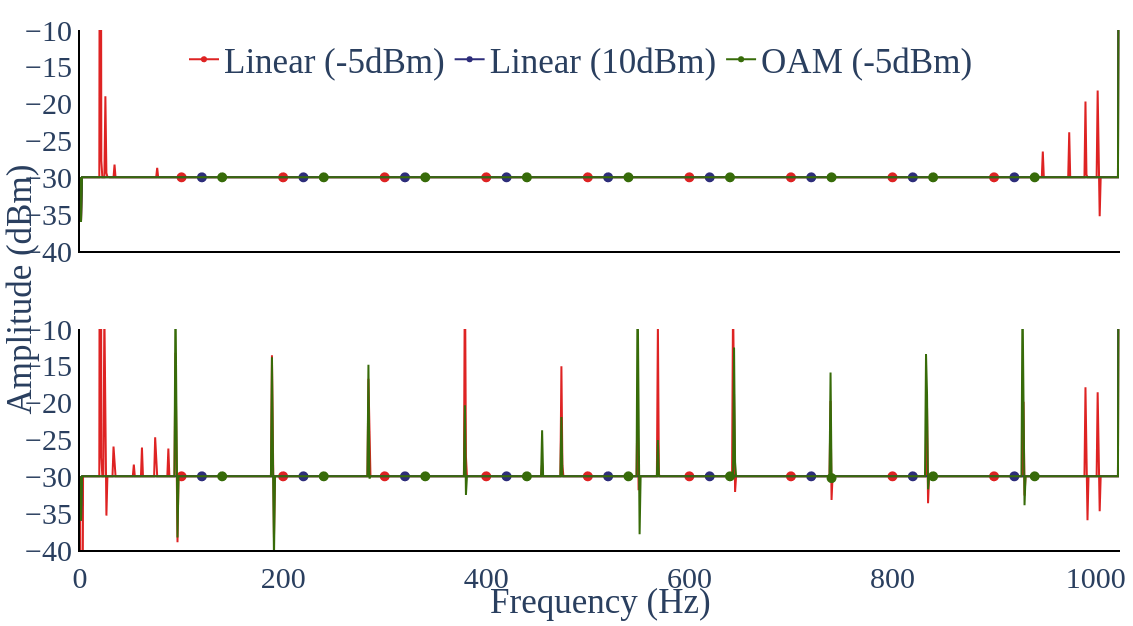}
\caption{(Top) Noise only and (Bottom) Signal with Noise frequency responses for linearly and OAM polarized antennas with different thresholds}
\label{fig:linearvsoam}
\end{center}
\end{figure}

Fig.~\ref{fig:linearvsoam} shows a comparison between the performance of linear and OAM antennas in an office environment with and without the presence of a rotating machine. The figures show the output of fast Fourier transform (FFT) at two different threshold values.  Fig.~\ref{fig:linearvsoam}(Top) shows the performance comparison in the presence of background noise only. The figure shows that the linear antenna receives a significant background noise, especially in the band below 100 Hz when the receiver threshold is -5 dBm. 
The receiver threshold needs to be increased to 10 dBm to suppress this background noise. Whereas with OAM antennas, the background noise is entirely suppressed at a receiver threshold of -5 dBm. Fig.~\ref{fig:linearvsoam}(bottom) show the results of FFT at various threshold values in the presence of a rotating machine.
The -5 dBm threshold level for the linear antenna causes interference from the noise to the actual machine rotation speed frequency that leads to the wrong estimation of the machine rotation speed frequency (will be shown in Sec.~\ref{sec:microbench}). When the receiver threshold is increased to 10dBm for linearly polarized antennas, the signal and noise are suppressed. Whereas the received signal for OAM waves with a receiver threshold of -5 dBm shows only the signal with minimal background noise, allowing for the design of lightweight signal processing algorithms.

This comparison between the received signals with linearly polarized and OAM waves show why \AlgoName uses specifically designed OAM antennas. We confirm the unreliability in the measurements made with linearly polarized antennas in Sec.~\ref{sec:evaluation}.

\section{System Overview}\label{systemoverview}

\begin{figure*}
\begin{center}
\includegraphics[width=0.75\textwidth,height=4cm]{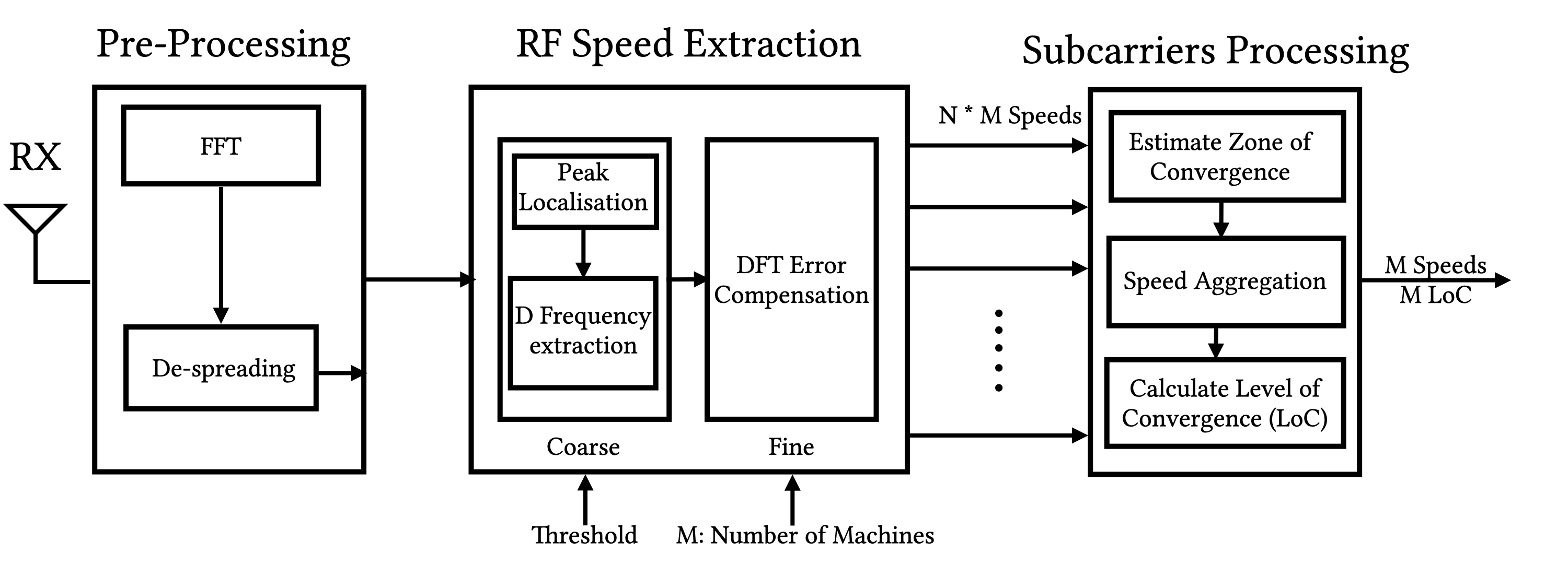}
\caption{\AlgoName receiver signal processing pipeline}
\label{fig:receiverflow}
\end{center}
\vspace{-4mm}
\end{figure*}

\AlgoName is a non-intrusive solution for measuring the rotation speed of multiple machines simultaneously.
It is a sensing system that consists of a transceiver capable of transmitting and receiving OAM electromagnetic waves and novel signal processing algorithms, dealing with real-world challenges.
\AlgoName encapsulates a monostatic radar sensing system. The TX and RX antennas are co-located inside the same device, allowing for a small form-factor that can be easily deployed in industries.
OAM antennas have two advantages: 1) higher immunity against depolarization in high scattering environment\cite{golmohamadi2016depolarization}, 2) OAM waves interact with the rotational motion by changing its frequency through rotational Doppler shift, which can be exploited to measure a machine's rotation speed.

The transceiver's frequency theoretically decides the minimum size of rotors that \AlgoName can measure. \AlgoName is currently designed to operate in the 5 GHz band, which can measure rotor sizes of up to 5.5 cm. However, \AlgoName's principles allow it to operate freely on other frequency bands. We chose to operate in the 5 GHz band due to its license exemption for radar-based sensing systems. This band is also shared with WiFi. However, WiFi is a broadband communication system designed with dynamic frequency selection in 5 GHz. Hence, if any WiFi device senses any narrowband communication in the band, it will change its frequency. 

\AlgoName starts by transmitting multiple narrowband signals on several subcarriers, where each subcarrier occupies a bandwidth of 1 kHz. The total occupied bandwidth by the transmitted signal is $N$ kHz for $N$ subcarriers. The $N$ subcarriers are divided across a 3 MHz bandwidth to avoid narrowband interference and comply with transmission power constraints. As a result, each 1 MHz has $N/3$ subcarriers that occupy $N/3$ kHz with an overall maximum transmission power of 50 mW.  

After the OAM waves are received, they undergo three distinct stages as shown in Fig. \ref{fig:receiverflow}: 

\fakeparagraph{1. Pre-processing:} 
The received scattered signal from the rotating machine can be expressed as per \cite{zheng2018analysis,chen2006micro}. Let $s_n(t)$ represent the scattered signal at the $n$-th subcarrier and $S(t)$ is the total scattered signal for all subcarriers. Also, let $\Gamma$ be the machine reflection coefficient magnitude and $\omega$ be the machine rotation speed. $s_n(t)$ can therefore be formulated as:
\begin{equation}
\begin{split}
    s_n(t) &= \Gamma_n \exp{\{jl\omega t\}}\exp{\{-jk_n\sqrt{R^2+{D^2}_z}\}}
    \\&\exp{\{-jk_n\sqrt{R^2+{D^2}_z-2Rd_rcos(\omega t)+{d^2}_r}\}}+z(n)
\end{split}
\label{equation3}
\end{equation}
\begin{equation}
    S(t) = \Sigma_{n=1}^{N}s_n(t)
\label{equation4}
\end{equation}
where $k_n$ is the wave vector magnitude of the $n$-th subcarrier. $z(n)$ is the additive white Gaussian noise with zero mean and $\sigma^2_z$ variance. Further, $R$ and $D_z$ represent the radial and z-axis position difference between the transceiver and the machine in cylindrical coordinates. $d_r$ is the distance between \AlgoName's transmitter and receiver. Eq.\ref{equation3} demonstrates how the rotation motion frequency modulates the carrier signal phase. This modulation process justifies the presence of the $\mathscr{D}$-frequency and its harmonics in the $S(t)$ frequency spectrum.
In this stage, the received signal is down-converted from 5GHz, an FFT analysis is conducted, and the signal is de-spread. The despreading block gathers the divided signal across 3 MHz as previously discussed into a single subband to ease processing.

\fakeparagraph{2. RF Speed Extraction (Sec.~\ref{speedextraction}):} The despread signal is fed as an input to the RF Speed extraction. The RF Speed extraction stage extracts the speed measurements of multiple machines from every subcarrier. This stage has two user-defined inputs, the number of machines to sense ($M$) and the minimum received signal threshold value. The RF Speed extraction stage is further divided into a coarse and fine speed estimation. First, the coarse estimator runs a peak localization algorithm to separate the signal from the noise. It then runs a $\mathscr{D}$-frequency extraction algorithm that leverages the machines' $\mathscr{D}$-frequency and harmonics to extract the speed measurements. The output of the coarse estimator is then used for the fine speed estimator. The fine speed estimator tunes the extracted speed measurements using a DFT error compensation algorithm. This fine speed estimator is one of the primary reasons for \AlgoName's 1 rpm resolution. The output of the RF Speed extraction stage is $N*M$, which feeds as an input to the subcarriers processing stage.

\fakeparagraph{3. RF Subcarrier Processing (Sec.~\ref{multiplesubcarriers}):} The final stage of \AlgoName's signal processing pipeline is RF subcarrier processing. \AlgoName is capable of operating in high-scattering environments and environments with RF noise. To deal with the problem of narrowband interference caused by dynamic human and object movements, we measure the speed on multiple frequency bands. We then aggregate these values and output the estimated speed for multiple machines and the level of convergence (LoC). The level of convergence is an indicator to the agreement between the majority of the subcarriers on a machine's rotation speed. 

\section{\AlgoName Speed Extraction}
\label{speedextraction}

In this section we describe how \AlgoName estimates the rotation speed of a single or multiple machines using a single subcarrier. \AlgoName addresses three challenges: It extracts the $\mathscr{D}$-frequency in the presence of multiple harmonics. It identifies the $\mathscr{D}$ individual frequency and harmonics of each machine in \AlgoName's field of view. Finally, \AlgoName estimates every machine's rotational speed. \AlgoName can measure a minimum rotation speed of 1 rpm, or $1/60$ Hz, and a maximum rotation speed of 7000 rpm with a sample time of $1$ second. \AlgoName's speed extraction algorithm is divided into two parts, first it does coarse speed estimation and then uses that estimation to do further fine speed estimation.

\subsection{Coarse Speed Estimation}
\label{coarsespeed}

\AlgoName first does a coarse rotational speed estimation using a Discrete Fourier Transform (DFT) to map the received signal's frequency response to frequency bins separated by $\Delta f = 1/T_d$, where $T_d$ is the sample window time. \AlgoName uses a sample time of $T_d = 1 s$, which means that $\Delta f = 1 Hz$. 
The received time series signal has frequency components that are non-negative real numbers, i.e. 100.2 Hz. When a DFT is run on the time series signal the frequency components are converted to whole(integer) numbers and lose the information in the decimal part. The non-integer(real) frequency component of the received signal is not really lost, it is distributed among the closest upper and lower frequency bins of the DFT spectrum output. For example, if the signal has a frequency component at 100.8 Hz, the DFT will output the maximum peak at 101 Hz and the lower peak at 100 Hz. For a received frequency component at 100.2 Hz, the DFT will output the maximum peak at 100 Hz and the lower peak at 101 Hz. The way that the DFT distributes the frequencies among the frequency bins is presented in detail in Subsec.\ref{finespeed}.
The DFT creates an error when it maps non-integer frequency components to integer frequency bins. This error propagates to the mapping of all of its harmonics as well. For instance, the frequencies 
100.8 Hz ($\mathscr{D}$-frequency), 201.6 Hz ($1^{st}$ harmonic), and 302.4 Hz ($2^{nd}$ harmonic) are mapped as 101 Hz, 202 Hz and 302 Hz respectively.

The scattered signal from the rotating machine being measured by \AlgoName contains the  $\mathscr{D}$-frequency and its harmonics. The scattered signal suffers from frequency selective fading that attenuates specific harmonic frequencies. For example, the received scattered signal might only contain the $\mathscr{D}$-frequency with the $3^{rd}$,$5^{th}$ \& $6^{th}$ harmonics, while the $2{nd}$ \& $4^{th}$ harmonics are completely attenuated. The loss of harmonics is a loss of information which makes it harder to find the true $\mathscr{D}$-frequency. To address the challenge posed by the loss of harmonics, \AlgoName finds the $\mathscr{D}$-frequency by calculating the greatest common factor (GCF) of the harmonics that are present in the received scattered signal.
What makes this part of \AlgoName a coarse estimation is the fact that the DFT creates an error when assigning the non-integer frequency components to integer values. The DFT introduces an error margin of $\pm 1 $ Hz. In our previous example the DFT gives us 101 Hz, 202 Hz and 302 Hz. 302 Hz is considered as the $2^{nd}$ harmonic of 101 because it lies in the range of 303 Hz $\pm 1$ Hz. It is clear that 101 Hz is the GCF of 202 Hz $\pm 1$ Hz and 302 Hz $\pm 1$ Hz, and the $\mathscr{D}$-frequency of rotation.

\AlgoName uses the same process to provide a coarse speed estimation for multiple machines. This is possible because we assume that each individual machine will have a distinct speed and will not share the $\mathscr{D}$-frequency or harmonics in the frequency domain.
\begin{enumerate}[noitemsep,leftmargin=\parindent,topsep=1.5pt]
    \item We start with the lowest frequency component and compare it against its following higher frequency components to determine the GCF between them.
    \item Eliminate the frequency components of the first machine.
   \item Repeat the first step for the next machine with the next-lowest speed of rotation.
\end{enumerate}

\subsection{Fine Speed Estimation}
\label{finespeed}
\AlgoName uses the output of the coarse speed estimator as input for the fine speed estimator. The fine speed estimator compensates for the error created by the DFT calculation and enhances our speed estimation to an accuracy of 1 rpm. 

We explain how \AlgoName does fine speed estimation by deriving how DFT spreads the amplitude of the non-integer frequency components among the closest neighbour frequency bins\cite{luo2016generalization}. 
We alter the derivation in \cite{luo2016generalization} to illustrate how this error decays far from the non-integer frequency components and better estimate the weight associated with each frequency bin. The altered derivation helps us in the implementation of fine speed estimation.

We first estimate the frequency response of the signal using continuous Fourier transform (FT) to estimate the error from the DFT. The signal has a frequency of $f_s$ and a duration of $T_d$. The observation time is such that $f_s >> 1/T_d$. The estimated frequency component $y(f_r)$  of the received signal $x(t) = exp(j2\pi f_st)$ can be calculated using FT as:
\begin{equation}
    y(f_r) = \int_{0}^{T_d}exp(j2\pi f_st)exp(-j2\pi f_rt)dt
    \label{eqn1}
\end{equation}
\begin{equation}
     y(f_r) = \frac{1}{j2\pi (f_s-f_r)}[exp(j2\pi (f_s-f_r)T_d)-1]
     \label{eqn2}
\end{equation}
Let $f_s-f_r = \frac{m}{T_d}$, where $m$ is real number, then Eq.\ref{eqn2} can be rewritten as:
\begin{equation}
    y(f_r) = \frac{T_d}{j2\pi m}[exp(j2\pi m)-1]
    \label{eqn3}
\end{equation}
Eq.\ref{eqn3} has different values according to $m$ which can be stated as:
\begin{equation}
y(f_r) = 
    \begin{cases}
        &(1) \hspace{10pt} T_d\hspace{110pt}m = 0\\
        &(2) \hspace{10pt} \frac{T_d}{j2\pi m}[exp(j2\pi m)-1]\hspace{35pt}0<m<1\\
        &(3) \hspace{10pt} 0 \hspace{113pt} m = n, \text{where } n\\
        &\hspace{135pt}\text{ is integer} \geq 1\\
        &(4) \hspace{10pt} \frac{T_d}{j2\pi m}[exp(j2\pi (m-n))-1]\hspace{12pt} n<m<n+1
    \end{cases}
\label{eqn4}
\end{equation}
Eq.\ref{eqn4} shows that the maximum amplitude is achieved at $m=0$. If the received signal frequency $f_s$ is a non-integer value, then $m$ is a non-integer value. This implies that $y(f_r)$ in Eq.\ref{eqn4} is either the second or the fourth case. 
DFT will be reflected in the frequency bins, $f_c \& f_f$, the two neighbour integer bins for the received non-integer signal frequency $f_s$. (\textit{i.e.} $f_c$ \& $f_f$ such that $f_c < f_s < f_f$, $f_c$ \& $f_f$ are two consecutive integer frequency bins). 
The second case indicates that the amplitude of the neighbour integer frequency bin ($f_c$ or $f_f$) is inversely proportional to its separation from $f_s$. 

The fine speed estimation algorithm uses the second case of Eq.\ref{eqn4} to calculate the non-integer(real) part of the received signal. The coarse frequency estimation in Subsec.~\ref{coarsespeed} successfully predicts the closest neighbour integer frequency component $f_c$ to the fundamental frequency $f_s$ (our estimated $\mathscr{D}$-frequency). The neighbour frequency bin ($f_f$) to $f_s$ should always be $f_c\pm1$. The amplitude of $f_f$ should be higher than other neighbouring frequency bins. \AlgoName compares the amplitudes of $f_c+1$ and $f_c-1$ and selects the highest as $f_f$. Substituting the $f_c$ and $f_f$ amplitudes in the second case of Eq.\ref{eqn4} gives the non-integer fundamental frequency component of the signal.

The received signal is accompanied by background noise, as illustrated in Eq.\ref{equation3}. The noise impacts the received signal-to-noise (SNR) level and can lead to an incorrect fine estimation of the $\mathscr{D}$-frequency. \AlgoName solves this by setting the minimum \textit{received signal threshold value} (discussed in Sec 3.2) that controls the permissible noise level in the receiver. As the threshold is increased more noise is filtered. The effect of the background noise and threshold level filtering are evaluated in Study 6 in Section \ref{sec:evaluation}.

Once \AlgoName has extracted the speed of multiple machines in each subcarrier, there are uncertainties in single subbands because of noise on the channel. \AlgoName processes readings from all subcarriers together to present the best speed estimate and the level of convergence, our metric that defines how many carriers are converging on similar values.

\section{\AlgoName subcarrier processing}\label{multiplesubcarriers}

This section describes our proposed multiple subcarrier algorithm that aims to mitigate the effect of narrowband noise on the scattered signals. Two important noise types that exist in the industrial environment are background additive white Gaussian noise (AWGN) and narrowband interference\cite{cheffena2016propagation}. The former type results from high operating temperatures, heavy machinery and a generally high-scattering environment. The latter type results from the dynamic change in the industrial environment, such as the movement of objects or humans in \AlgoName's field of view. The AWGN is spread over the entire RF sensing spectrum; however, the narrowband interference is concentrated on specific bands.\AlgoName's multiple subcarrier algorithm addresses the challenge of interference from both types of noise.

In the multiple subcarrier algorithm, the transmitted power is spread over $N$ subcarriers distributed over a 3 MHz bandwidth. At \AlgoName's receiver, each subcarrier is passed through the speed extraction algorithm, discussed in Sec.\ref{speedextraction} to estimate the machine rotation speed with 1 rpm resolution. First, the multiple subcarrier algorithm defines the \textit{Zone of Convergence (ZoC)}, which is defined as the range that encompasses the rotation speed estimated values for most of the subcarriers. As previously mentioned in Sec.\ref{speedextraction}, \AlgoName's speed extraction algorithm relies on amplitude comparison between the two nearest neighbour integer frequencies $f_c$ and $f_f$ to determine $f_s$. Since the amplitude can differ from one subcarrier to another because of the difference in magnitude of the random background noise, any subcarrier's estimated $\mathscr{D}$-frequency should lie within $\pm 1$ Hz from a centre or average value. Consequently, the ZoC includes the range of estimated rotation speeds from all subcarriers that lie within is $\pm 60$ rpm from a median rpm value.

After defining the ZoC, the multiple subcarrier algorithm aggregates only the estimated speeds from the subcarriers inside the ZoC and considers other subcarriers' estimated speeds as outliers. The aggregation process takes the average value of the estimated speeds from the subcarriers in the ZoC. The number of subcarriers that give speed estimation within ZoC range represents the \textit{LoC}. The ratio of LoC to the total number of subcarriers $N$ is defined as the LoC ratio. If the LoC ratio is high (\textit{i.e. > 95 \%}), this means that more than 95\% of the subcarriers give speed estimation with a maximum standard deviation of $\pm 1$ Hz. This high LoC ratio implies that the deviation in the speed estimation of each subcarrier can be mitigated through the aggregation process over multiple subcarriers and computing the average estimated speed. However, if the LoC ratio is below 95\%, this means that deviation in the estimated speed for each subcarrier by the background noise is high enough to create an outlier ratio $>$ 5\%. In the latter case, where LoC is $<$ 95\%, the threshold level and transmission powers should be varied to a level that accommodates the effect of the background noise and achieves a high LoC ratio.

Now we know that a high LoC ratio can guarantee reliable performance in the presence of AWGN background noise; however, this may be insufficient in the presence of narrowband noise. Therefore, the LoC magnitude is a good indicator of the system reliability in the presence of narrowband interference. High LoC magnitude means many subcarriers are used that are spread over the 3 MHz bandwidth. This spreading over large bandwidth mitigates the effect of narrowband interference. As a result, we require a high LoC ratio and a high LoC magnitude to mitigate both effects of background noise and narrowband interference. However, the multiple subcarrier algorithm spreads the transmission power among all the subcarriers, limiting the coverage distance if we use a high number of subcarriers. Hence, there is a trade-off for the \AlgoName between the performance reliability and coverage distance determined by the level of noise in the industrial environment.

\section{Implementation}
\label{sec:implementation}

In this section, we discuss the implementation of \AlgoName. 
We test our system on eight types of machines with different sizes, materials and line-of-sight availability. We also describe the implementation of our OAM antennas.

\subsection{Implementation Details}

\begin{figure}
\begin{subfigure}{0.26\textwidth}
  \centering
    \includegraphics[width=\textwidth]{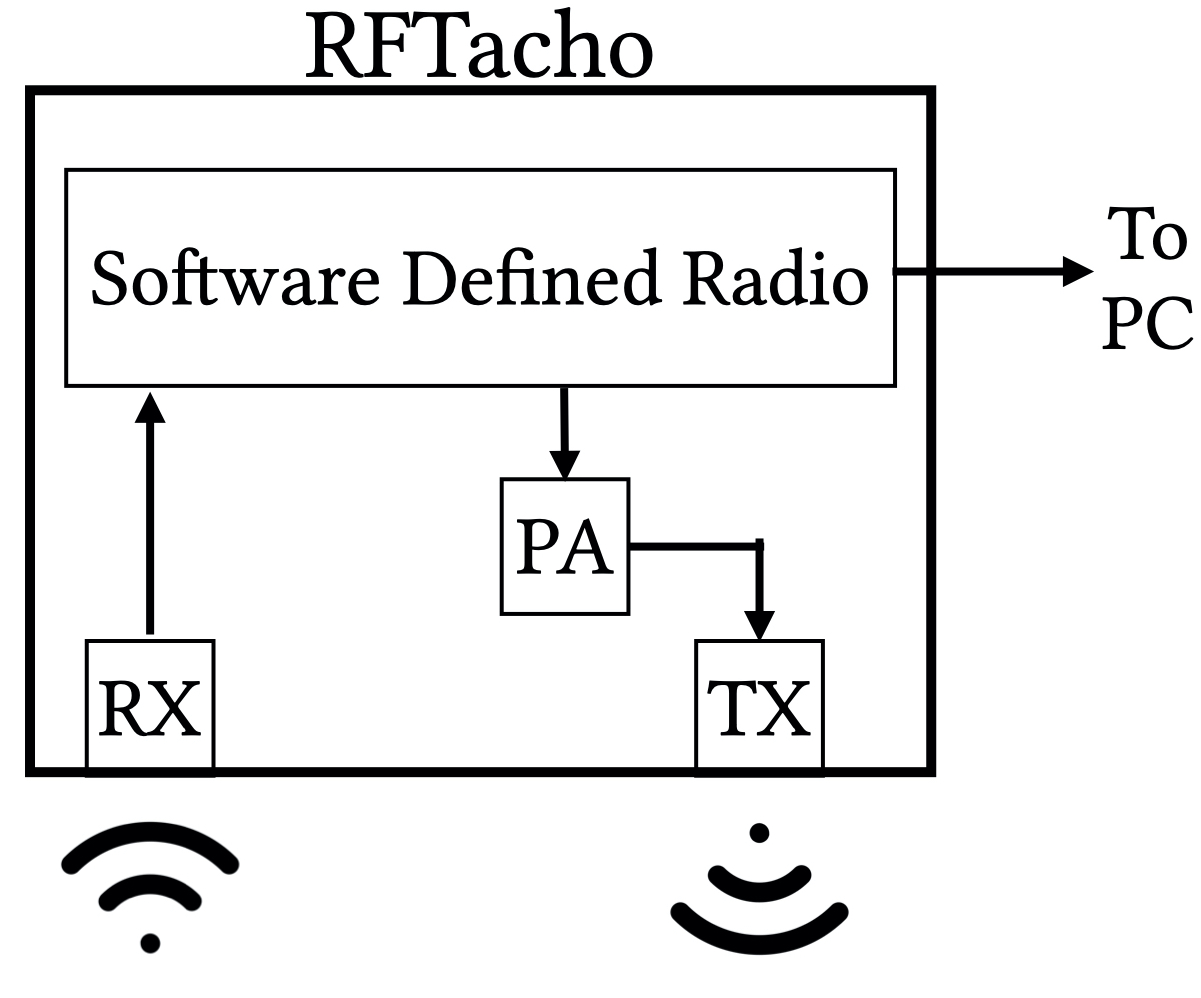}
    \caption{\label{fig:implementation_data}}
\end{subfigure}%
\begin{subfigure}{0.26\textwidth}
  \centering
    \includegraphics[height=3cm]{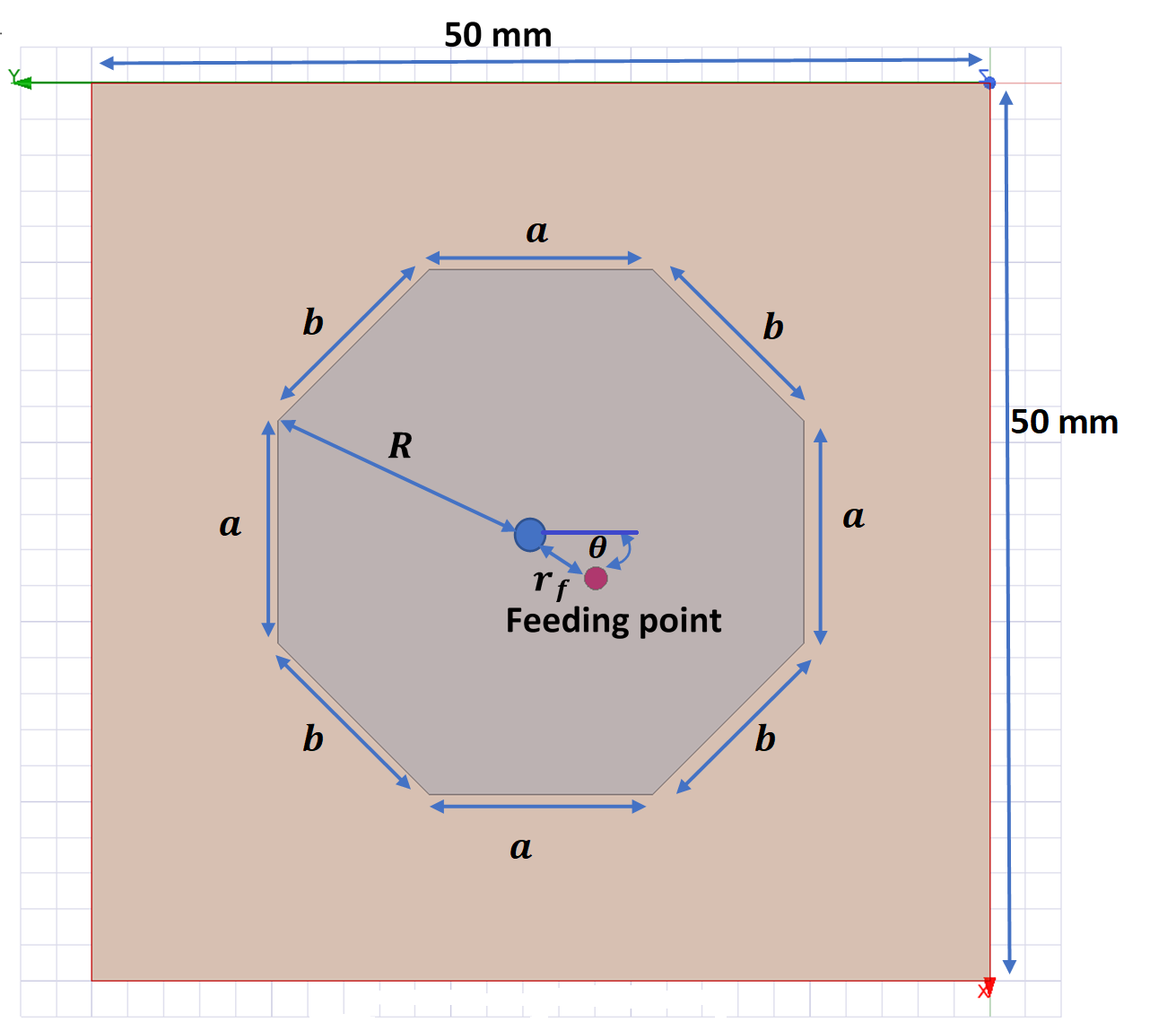}
    \caption{\label{fig:implementation_real}}
\end{subfigure}%

\caption{Implementation Setup (a) Functional diagram (b) Octagon OAM patch antenna design}
\label{fig:implementation_setup}
\vspace{-5mm}
\end{figure}

\AlgoName is a hardware-software sensing system. The antenna implementation for \AlgoName is defined in Sec.~\ref{sec:antennadesign}.
We use a USRP B210 Software Defined Radio (SDR) with two antennas connected to Port 1. The first antenna is the transmitter, and the second is the receiver. The USRP B210 has a maximum sampling rate of 61.44 Msps, and a maximum transmit gain of 10dB. An external power amplifier, QPA9501, is used at the transmit antenna to boost the transmission power up to 22dBm ($\sim150 mW$). The SDR communicates over USB 3.0 to a PC running GNU Radio 3.7. The implementation setup and the hardware antenna design are shown in Fig~\ref{fig:implementation_setup}.

\subsection{OAM Antenna Details}
\label{sec:antennadesign}

We design our OAM antenna based on \citeauthor{guo2019oam} \cite{guo2019oam}, but it has a smaller area compared to their design and operates at 5.525 GHz. This compact antenna size facilitates the deployment of our monostatic radar sensing system in an industrial environment. 

We propose an octagon patch antenna. Our proposed design antenna is shown in Fig.~\ref{fig:implementation_real}. The dimensions of the antenna are calculated using the ANSYS HFSS software package, such that the antenna input impedance is 50 $\Omega$ at the desired operating frequency. The radius of the octagon $R$ is 15.9 mm, and the dimensions of the sides are $a=12.4mm$ and $b= 11.7 mm$. The feeding point of the antenna is placed at distance $r_f$ = 2.6 mm and angle $theta = 40^{\circ}$ from the octagon centre. The octagon is placed over an electric conductor ground plane. The dimensions of the ground plane are $50 mm \times 50 mm$. The dielectric material between the octagon and the ground plane is \textit{FR-4}, which has a relative permittivity of 4.2 and a thickness of 1 mm. The overall antenna dimensions are $50 mm \times 50 mm$, which is nearly one-ninth of the antenna proposed in \cite{guo2019oam}.

\subsection{Choosing the transmission gains}
\label{subsec:transmitgain}

\AlgoName operates in the 5.47-5.73 GHz band, defined in Band B of the WiFi regulations\cite{ofcom}. Devices operating in this band can operate indoors and outdoors. The devices can have a maximum E.R.P. of 1 Watt and a maximum PSD of 50 milliWatt/MHz in any 1 MHz band.
\AlgoName needs to adhere to the regulations, and so it distributes the subcarriers in different 1 MHz channels. The maximum number of subcarriers per channel, the number of channels and the SDR gain are chosen to ensure the maximum PSD is not exceeded. The maximum E.R.P. is always under 1 Watt in our implementation. The results for maximum gain settings and the number of channels are shown in Table~\ref{tab:transmissiongain}.

\begin{table}[]
	\caption{Transmission parameters}
	\label{tab:transmissiongain}
	\resizebox{\columnwidth}{!}{
	\begin{tabular}{|c|c|c|c|}
	\hline
	\textbf{Subcarriers (N)} & \textbf{Max Transmit Gain (dB)} & \textbf{Number of channels} & \textbf{E.R.P. (mW)} \\
	\hline
		 1 & 63 & 1 & 50 \\
		 5 & 76 & 5 & 150 \\
		 15 & 79 & 3 & 150 \\
		 30 & 85 & 3 & 150 \\
		 60 & 90 & 3 & 120 \\
	\hline
	\end{tabular}}
\vspace{-5mm}
\end{table}

\subsection{Test machines}
\label{subsec:machines}

We test \AlgoName on eight machines. The machines have been chosen to test the different speeds, materials and diameters for which \AlgoName works reliably. The machines and their properties are:
\begin{enumerate}[leftmargin=10pt,topsep=1.5pt]
    \item \textbf{USB-powered fan (2x):} We use two fans with different materials, sizes and number of blades. The first fan has four \textit{metallic blades}. The radius of the blade is 5.2 cm. The second fan has three \textit{Polypropylene} blades. The radius of each blade is 9 cm. Both fans do not have a feedback loop, so the speed is not constant and changes depending on the current. 
    \item \textbf{Axial fan:} An axial fan used in computers case for active cooling. The fan has five plastic blades with a diameter of 7.1 cm. The axial fan's speed can be varied depending on the input voltage. The speed varies from 3500-7000 rpm when the voltage is varied.
    \item \textbf{Hand drill:} A hand drill with a metallic chuck diameter of 10mm. The drill has a pressure-controlled speed with a maximum speed of 1400 rpm. 
    \item \textbf{Brushless DC motor (BLDC)\cite{bldcmotor}:} A BLDC motor, a widely used component in drones, electric vehicles, industrial robots and several other industries. We use the 9225-160KV Turnigy Multistar BLDC motor, which has a maximum speed of 4500 rpm.
    \item \textbf{Two phase oil and water separator\cite{breza2020separator}:} An industrial process testbed designed to replicate the oil and water separation process in the oil industry. The testbed consists of a variable speed 4-pole motor. The motor has a maximum safe operating speed of 1800 rpm with a shaft diameter of 4cm. 
    \item \textbf{Ceiling Air-conditioner:} A ceiling air-conditioner (AC). The AC has an embedded fan that is completely covered by the a metallic ventilation grille. The grille has horizontal slots to allow the air flow. The AC has a DC motor to rotate the fan with a maximum speed of 1800 rpm.
    \item \textbf{Washing machine:} A household washing machine. The machine has an internal metallic drum of a radius of 19 cm and a controlled operation speed of a maximum of 1600 rpm.
\end{enumerate}

\section{Evaluation}\label{sec:evaluation}

This section will show \AlgoName's capabilities of: measuring the speeds of multiple machines, operation in high-scattering environments, speed span, impact of interference and finally, the speed outputs of linear and OAM antennas.

\fakeparagraph{Study 1: Simultaneous speed measurement for multiple machines}
In the first study, we showcase \AlgoName's multiple machines speed measurement capability. For this study, we run three machines from our list of test machines. Two of these machines are USB-powered fans, and the third machine is the voltage controlled axial fan. The experiment is performed with $10$ subcarriers. The distance between \AlgoName and the three machines is $30 cm$.

\begin{figure}
\includegraphics[width=0.45\textwidth, height = 3.5cm]{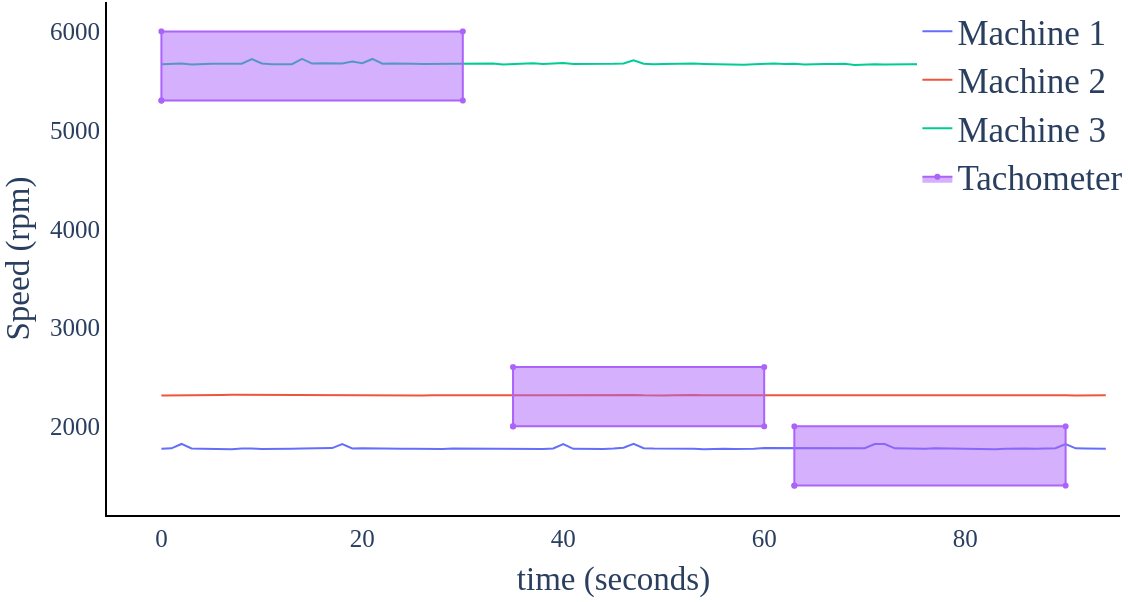}
\caption{\AlgoName simultaneous speed estimation of multiple machines vs. tachometer single machine speed estimation at a time}
\label{fig:multiplemachine}
\vspace{-5mm}
\end{figure}

\textbf{Experiment Results.} Fig.~\ref{fig:multiplemachine} shows the speed estimation for the three machines using \AlgoName and the tachometer. The results show that the error between \AlgoName and tachometer is less than $0.5\%$. However, a tachometer can only measure the speed of a single machine at a time, whereas \AlgoName works reliably for all machines. This study clearly shows how beneficial \AlgoName is compared to traditional sensors.

\fakeparagraph{Study 2: Speed measurement of single machine.}
In the second study, we will test the stability of the \AlgoName speed estimation algorithm using multiple subcarriers and
in the presence of a single rotating machine. The machine used in this experiment is the USB fan. The number of subcarriers used is $60$, and the distance between the \AlgoName and the fan is 22 cm.

\begin{figure}
\begin{center}
\includegraphics[width=0.47\textwidth, height = 3.5cm]{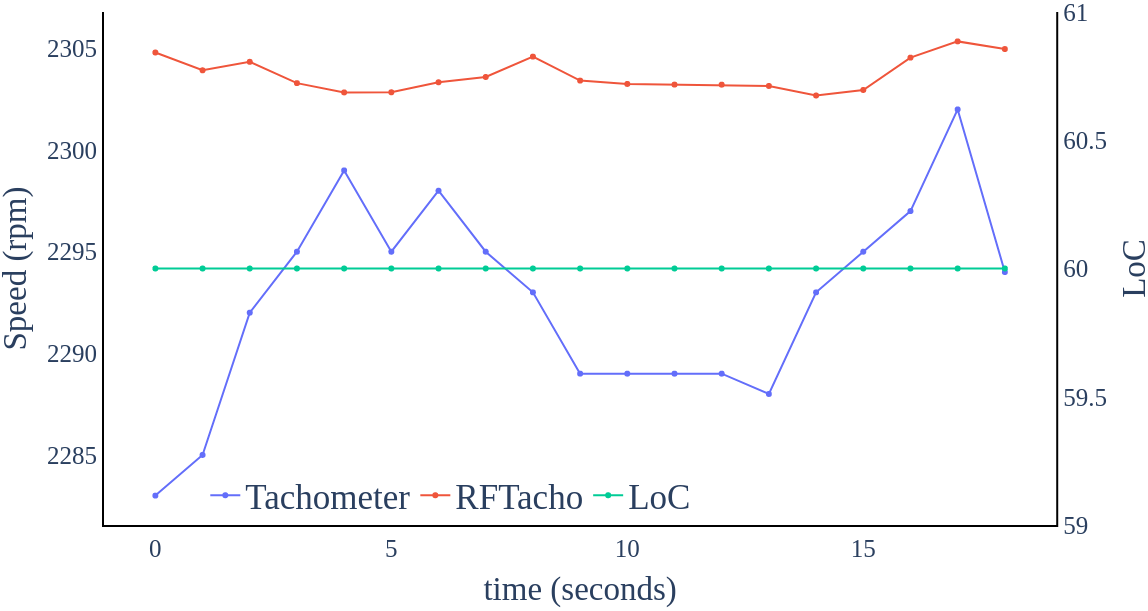}
\caption{Single machine}
\label{fig:singlemachine}
\end{center}
\vspace{-5mm}
\end{figure}

\textbf{Experiment Results.} Fig.~\ref{fig:singlemachine} shows the speed estimation of \AlgoName versus the tachometer speed measurement. The left vertical axis shows the rotation speed values, and the right axis shows the LoC values. The figure shows the stability in the reading of \AlgoName for the rotation speed using $30$ subcarriers, where the average value is 2303 rpm, and the standard deviation is $\pm 1 rpm$. Moreover, the tachometer reading shows an average of 2293 rpm and a standard deviation of 5 rpm. The LoC is 60, which means that all the subcarriers' speed estimations lie in the same ZoC and converge to the same average reading, which justifies the high stability in \AlgoName reading compared to tachometer reading.

\fakeparagraph{Study 3: OAM vs. linearly polarized antennas.} In the third study, we compare the performance of \AlgoName in the presence of OAM antennas and linearly polarized antennas while reporting on the LoC. We use $10$ subcarriers for both antennas sets and the voltage-controlled axial speed fan. The voltage is set to 19 V, which corresponds to an output speed of 5676 rpm. The distance between the fan and the \AlgoName is $30 cm$.

\textbf{Experiment Results.} As early shown in Fig. \ref{fig:linearvsoam}, and discussed in Subsec.~\ref{antennacomparison}, linearly polarized antennas suffer from severe depolarization in high scattering environments. This problem reflects on noise interference in the lower frequency bands and leads to poor speed estimation. Fig. \ref{fig:linearvsoam} shows the speed estimation of both sets of antennas in two cases: 1) the presence of the background noise only, and 2) in the presence of the background noise and a rotating machine. In the first case, the linearly polarized antenna cannot discard the interfering noise at a minimum input signal threshold of -5 dBm, which leads to the wrong speed estimation of 1838 rpm. This wrong estimation has a high standard deviation of 239 rpm and a low LoC of 6.71 (<95\%). In the presence of the rotating machine, the linear antenna leads to a speed estimation of 5458 rpm with a high standard deviation of 948 rpm and a low LoC of 8.83 (<95\%). If we increase the threshold to 10 dBm, the linearly polarized antenna can discard the noise. However, this threshold level fails to estimate the machine speed in the second case and gives 0 rpm output reading. Compared to linearly polarized antennas, the OAM antennas can discard the noise at -5 dBm threshold level and leads to an accurate speed estimation for the axial fan with a low standard deviation of 1 rpm and high LoC (100\%).

\fakeparagraph{Study 4: Fine speed estimation vs. coarse speed estimation.}
In the fourth study, 
we show the effectiveness of \AlgoName's speed estimation sampling at 1 second by comparing it against coarse speed estimation at different sampling time windows without a fine estimator. 
Longer sampling time windows can obtain signal frequency resolution $<1$ Hz, which means a lower rpm resolution. We use the USB-powered fan as our test machine. We measure the speed using the coarse speed estimation algorithm at a sampling time window that spans the range from 1 to 30 seconds.

\begin{figure}
\includegraphics[width=0.47\textwidth]{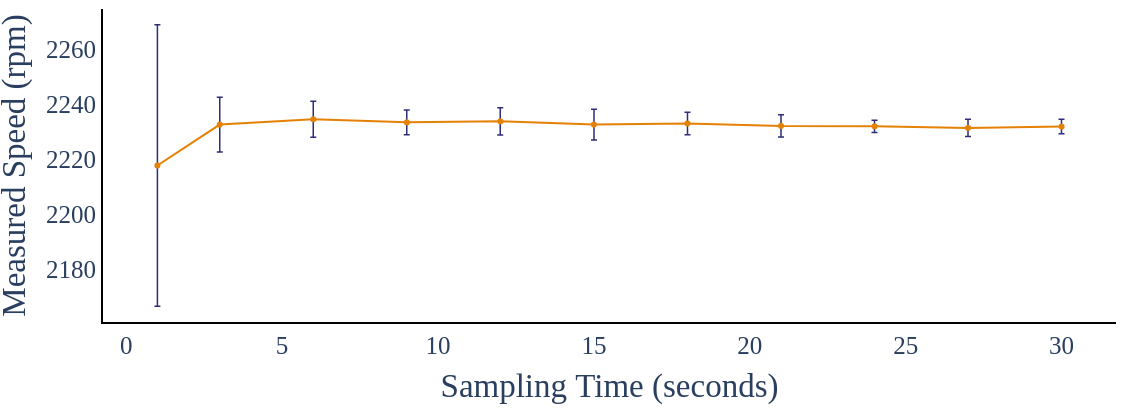}
\caption{\AlgoName coarse speed estimation (rpm) at different sampling time windows (seconds)}
\label{fig:samplingtime_result}
\vspace{-5mm}
\end{figure}

\textbf{Experiment Results.} Fig.~\ref{fig:samplingtime_result} shows the estimated speed based on the coarse speed estimation algorithm alone. The results show how the standard deviation of the speed estimation decreases with the increase in time. \AlgoName's speed measurement has a standard deviation of 4 rpm and a mean error of less than 0.8\% for a 1s sampling time. The coarse speed estimation algorithm alone shows a standard deviation of 4 rpm in 18s compared to 51 rpm in 1s. Moreover, the optical tachometer speed estimation has a standard deviation of 5.9 rpm. These results prove that \AlgoName's coarse estimator with the fine estimator provides stable readings in a sampling window of 1s which would take up to 18 seconds to obtain with a coarse estimator alone.

\fakeparagraph{Study 5: Measuring the impact of subcarriers on speed measurements and LoC in presence of narrowband noise span of \AlgoName.} In the fifth study of our evaluation, we study how \AlgoName performs in the presence of narrowband noise. We simulate narrowband noise by moving objects randomly between \AlgoName and the test machine. We run several experiments where we increase the number of subcarriers from 1 to 60 and report on the measured speed values and LoC.

\begin{figure}
\includegraphics[width=0.47\textwidth]{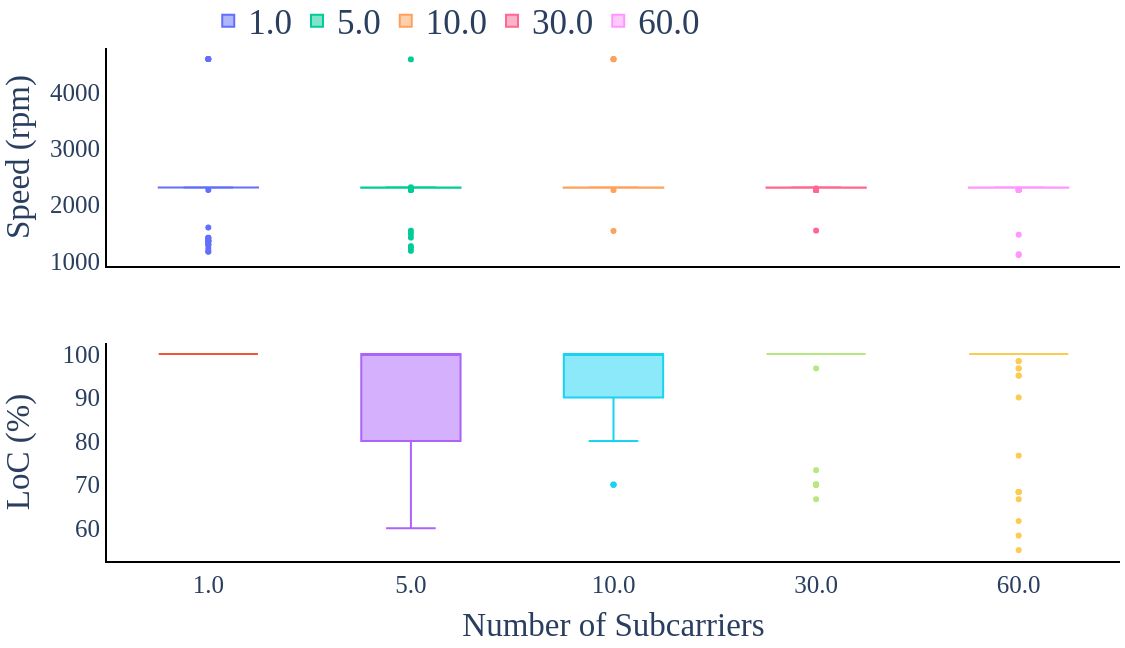}
\caption{Impact of number of subcarriers on speed measurements and LoC in presence of narrowband noise (Top) Speed Measurements (Bottom) LoC values}
\label{fig:noise_movements}
\vspace{-4mm}
\end{figure}

\textbf{Experimental Results.} Fig.~\ref{fig:noise_movements} shows the results of our fifth study. The results show that all subcarriers can estimate the speed of the rotating machines accurately in the majority of the scenarios. With a low number of subcarriers ($1$), the LoC values are always 100\%, so the system cannot distinguish between a correct and incorrect measurement. As we increase the number of subcarriers, we observe that the number of outliers reduces. The system can identify incorrect readings caused by narrowband noise by reporting on the LoC values.

\fakeparagraph{Study 6: Measuring the impact of narrowband interference and background noise on the speed estimation error.} In the final study, we evaluate the effect of narrowband interference and background noise on \AlgoName's accuracy at various received SNRs. For narrowband interference, the source of interference is another rotating machine in \AlgoName's measurement area. We simulate two cases of interference, where the interference source rotates at a similar speed ($\pm 10$ rpm) and at different speeds from our target rotating machine. In the first case, we use two stand fans (one as the interferor and the other as target). Both fans rotate at an average speed of 1227 rpm. In the second case, we use a stand fan (1227 rpm) as the interference and a USB-powered fan (average speed 2242 rpm) as the target. 

For background noise, we simulate the effect of the background noise on our speed estimation by combining \AlgoName's receiver with a additive white Gaussian noise (AWGN) generator with different powers. The AWGN generator is added to the received signal before the pre-processing block in Sec. 3.1. We conduct the experiment at two different signal threshold levels at the receiver: -20 and -5 dBm.

\begin{figure}
\includegraphics[width=0.47\textwidth]{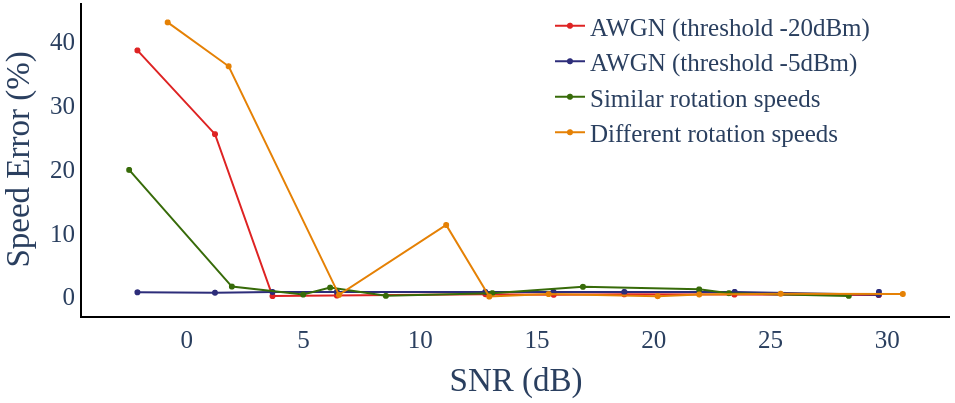}
\caption{Impact of noise interference and background noise on \AlgoName speed estimation error}
\label{fig:error_SNR}
\vspace{-4mm}
\end{figure}

\textbf{Experimental Results.} Fig.~\ref{fig:error_SNR} shows \AlgoName's speed estimation error. SNR is computed at the receiver and before the pre-processing block, and it automatically accounts for the noise type applied to the system. In the case where the interference source and target machines rotate at similar speeds, \AlgoName obtained an error $<1.5\%$ at SNR $>8.5$ dB. This is due to frequency selective fading that arises as a result of destructive interference from scattered signals of two sources rotating at the same speed and generating similar frequency reflections. In the second case where the interference source and rotating machine rotate at different speeds, \AlgoName achieves a speed estimation error  $<1.5\%$ at SNR $>13$ dB. This is due to the wrong detection of the interference source as the main target machine at SNR levels $<13$ dB. This problem is already solved in \AlgoName's multiple machine detection algorithm, evaluated in Study (1), where by setting $M=2$, we will be able to accurately measure the speed of both machines.

Fig.~\ref{fig:error_SNR} also shows the effect of AWGN on the speed estimation error. We evaluated the performance at two threshold levels at \AlgoName's receiver: -20 and -5 dBm. At -20 dBm threshold level, a high SNR is required at the receiver to obtain low estimation error. However, the results obtained by the -5 dBm threshold level show a reliable speed estimation error $<1\%$ at SNR $<0$ dB. The results show that \AlgoName maintains its high accuracy by filtering out background noise which is feasible by selecting the appropriate threshold levels.

This section showcased \AlgoName's capability of measuring the speed of multiple machines, its speed span, its resilience to noise and why we chose OAM antennas instead of linearly polarised ones. The results show the advantages of \AlgoName compared to optical tachometers and how \AlgoName can be easily deployed due to its small form factor.
\section{Micro-benchmarks}
\label{sec:microbench}

In this section, we will further demonstrate \AlgoName's sensing capabilities. We describe these as micro-benchmarks as they demonstrate the sensing capabilities like sensing range, orientation, non-line-of-sight operation and how it performs with machines of different sizes and rotor compositions.

We conduct four performance studies to show \AlgoName's performance in different situations.\AlgoName operates in the same wireless frequency as WiFi. We conduct all experiments by keeping the wireless environment as constant as possible and ensuring minimal external interference. The ground truth is measured with a commercial optical tachometer with a maximum measurement error of 0.5\% of the speed measurement. We manually vary the gain and threshold settings until the desired LoC is achieved and the results with the highest LoC are reported ($>95\%$).

\fakeparagraph{Study 1: Varying the distance between \AlgoName and the machine.} The first study we conduct measures the effect of changing the distance between \AlgoName and the machine. We use the USB-powered fan for this experiment, and the distance is measured from the centre of two antennas and the machine itself. The distance is varied as multiples of the wavelength (5.5 cm). We use $N=60$ for this experiment, giving us the least distance but the highest resolution (up to 1 rpm). 

\begin{figure}
\includegraphics[width=0.47\textwidth]{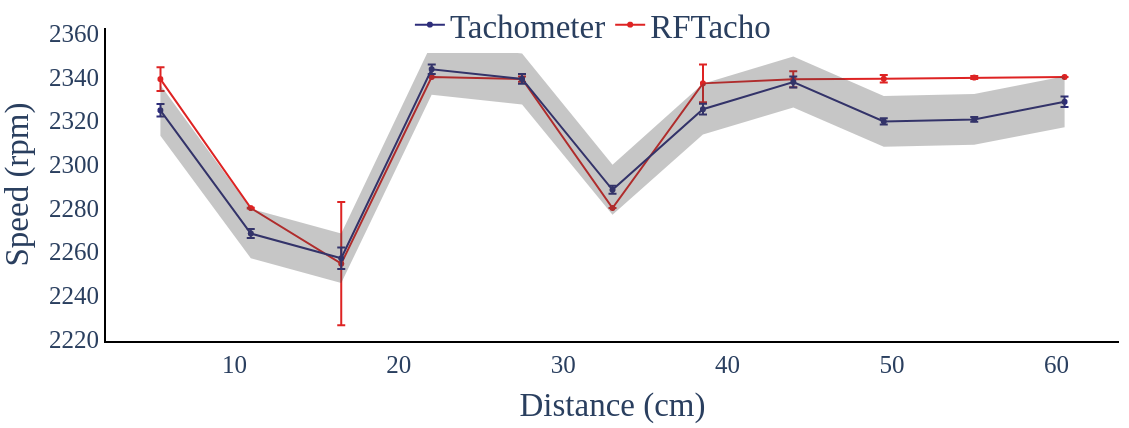}
\caption{Effect of distance}
\label{fig:distance_result}
\end{figure}

\textbf{Experiment Results.} The results for this experiment are shown in Fig.~\ref{fig:distance_result}. The tachometer has a maximum error of 0.5\% of the speed indicated in grey in the figure. The results show that the maximum mean error of \AlgoName is 0.8\% which is close to the maximum error of an optical tachometer. The results also show that the error is lower than 0.5\% at several distances. This study shows that \AlgoName, with this configuration, can operate up to 61.5 cm distance with 60 subcarriers and demonstrate low error margins compared to a state-of-art tachometer.

\fakeparagraph{Study 2: Effect of changing the orientation between \AlgoName and the machine.}
The second study examines how the orientation between \AlgoName and test machine affects the system's error. The optical tachometer always has to be head-on with the reflective tape of the machine, and any deviation from this leads to error or no readings. With this study, we are interested in measuring how \AlgoName performs when there is a tilt between machine and itself. 
The USB-powered fan is placed at an angle relative to the straight-line perpendicular to the antennas. We use 60 subcarriers and measure the maximum distance and orientation for which \AlgoName measures the machine's speed reliably.

\begin{table}[!ht]
	\caption{Effect of changing orientation}
	\label{tab:orientation}
	\centering
	\begin{tabular}{|c|c|c|}
	\hline
	\textbf{Orientation (degrees)} & \textbf{Max distance (cm)} & \textbf{Mean error (\%)} \\
	\hline
		 90 & 60 & 0.15 \\
		 45 & 60 & 0.4 \\
		 0 & 65 & 0.3 \\
		 -45 & 62.5 & 0.05 \\
		 -90 & 60.5 & 0.79 \\
	\hline
	\end{tabular}
	\vspace{-3mm}
\end{table}

\textbf{Experiment Results.} Table~\ref{tab:orientation} shows results of our study. The results show that \AlgoName works reliably from -90 to +90 degree orientations for up to 65 cm. The mean error in the scenarios is less than 0.8\% of the optical tachometers. The results show an important feature of \AlgoName where optical tachometers fail. \AlgoName works reliably even when it is not facing the rotating object head-on. This feature is useful when the sensor cannot be placed directly in front of the object to be sensed.

\fakeparagraph{Study 3: Effect of placing objects between \AlgoName and the machine recreating non-line-of-sight paths.} In the third experiment, we study the effect of the non-line-of-sight (NLOS) path between \AlgoName and the machine. To do this, we enclose the machine with four boxes made of different materials. The four objects (materials) are plexiglass of 1.1mm thickness, Styrofoam, cardboard and a transparent plastic box. We use the axial fan running at a constant speed, placed at a distance of 19 cm from \AlgoName operating with 60 subcarriers.

\begin{figure}
\includegraphics[width=0.47\textwidth]{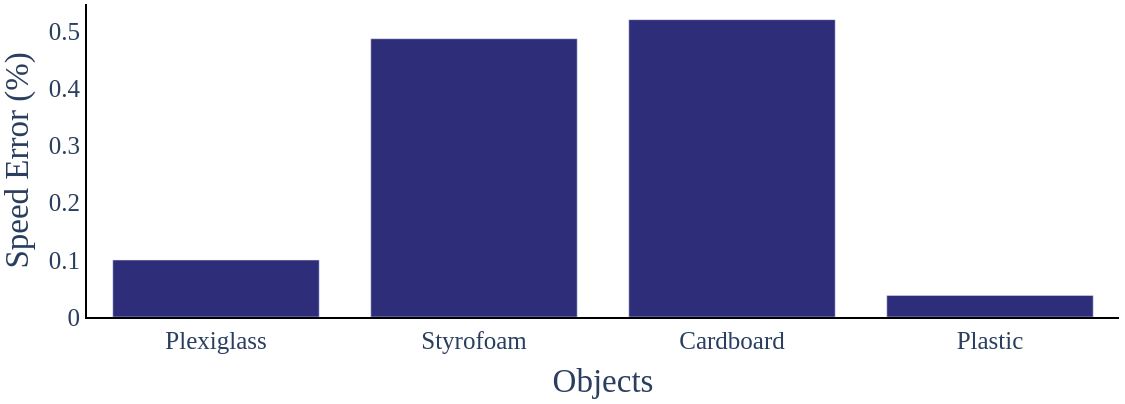}
\caption{Non-line-of-sight results error compared to a tachometer, speeds for Styrofoam and cardboard are estimated as tachometer does not work in these scenarios}
\label{fig:nlos_result_error}
\vspace{-4mm}
\end{figure}

\textbf{Experiment Results.} Fig.~\ref{fig:nlos_result_error} shows results of this study. The mean error for \AlgoName compared to the tachometer for plexiglass and the transparent plastic box are less than 0.1\%. The optical tachometer fails with the Styrofoam and cardboard as there is no line-of-sight to the machine. The mean error for \AlgoName, assuming the axial fan's speed, is less than 0.8\% with Styrofoam and cardboard. This experiment shows how \AlgoName works in scenarios where optical tachometers fail.

\fakeparagraph{Study 4: Diverse machine types}
In our final experiment, we study how \AlgoName performs with industrial-grade machines. For this experiment, we measure the speed of 5 machines, namely, a hand drill, two-phase separator, ceiling air-conditioner, washing machine and a BLDC motor. A detailed description of these machines is done in the Subsec.~\ref{subsec:machines}. For these experiments, the optical tachometer was used to measure the actual speeds in scenarios where it was feasible.

\begin{figure}
\includegraphics[width=0.47\textwidth]{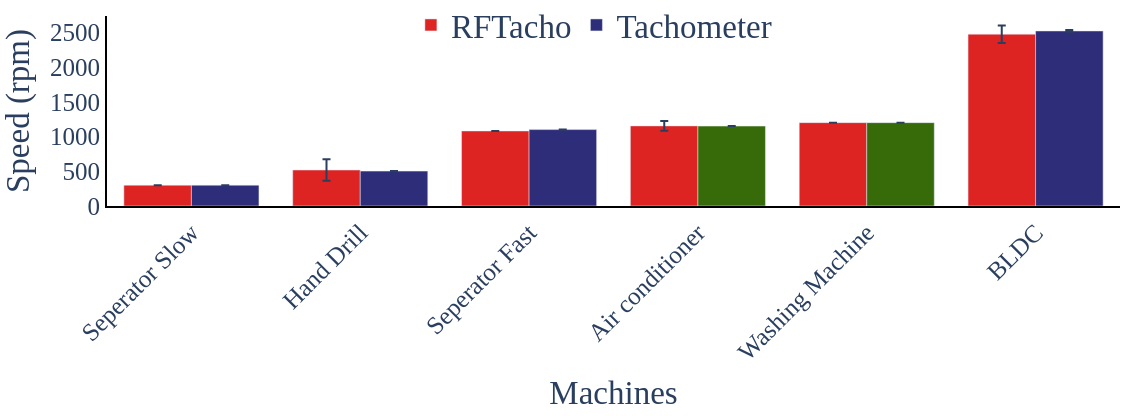}
\caption{Multiple machines results (Green indicates speed was estimated)}
\label{fig:multiple_machines_result}
\vspace{-4mm}
\end{figure}

\textbf{Experiment Results.} Fig.~\ref{fig:multiple_machines_result} shows the results of this experiment. The results show that \AlgoName has a maximum error of 3\% in the hand drill scenario, less than 2\% in the two-phase separator at multiple speeds and no error in the BLDC motor. The ground truths for the air conditioner and washing machine were hard to measure as the rotating parts are not easily exposed or measurable with a tachometer. However, we assume ground truths based on the specifications, and the error is less than 0.25\%. The results show that \AlgoName can work well in several industrial scenarios and even in scenarios where optical tachometers fail, like the air conditioner and washing machine. 

\textbf{Experimental discussions:}
The results show that \AlgoName is a viable substitute for tachometers and offers benefits over traditional sensors. \AlgoName works well with several machines and at various distances. Unlike traditional tachometers requiring a line-of-sight to the rotating object, \AlgoName works in non-line-of-sight situations and at different orientations. These features open up several use-cases for \AlgoName where a tachometer would fail, like if the rotating object is embodied in a cover. The results also show a maximum distance versus resolution trade-off that must be considered while deploying the system. Finally, the results show that \AlgoName works well in a high-scattering environment, like in a factory environment. Moreover, a study has been conducted using the axial fan to show the speed estimation span of \AlgoName. In this study, the voltage-controlled axial fan speed has been varied between 3800 rpm (12V) to 6700 rpm (24V). The results show that \AlgoName's speed estimation has a mean error of less than 0.6\% of the tachometer's speed in all scenarios.

\section{Related Work}
\label{sec:relatework}

\fakeparagraph{RF Sensing.}
RF sensing spans a wide range of applications as a non-invasive, low-cost and contactless monitoring solution. These applications include: human health care like breathing and heart rate monitoring\cite{shah2019rf},
target localisation in search and rescue missions \cite{wang2016rf}, 
automotive sensing applications \cite{zheng2020v2ifi}, temperature measurement \cite{chen2020thermowave}, acoustic signal identification from multiple sources and machine vibration monitoring \cite{wang2020uwhear}. However, the signal processing algorithms used in previous literature %
\cite{shah2019rf,wang2016rf,alanis20143d,zheng2020v2ifi,chen2020thermowave} are impractical to use for rotation speed monitoring of some machines. The vibration monitoring techniques used in \cite{wang2020uwhear} use RF signals to sense acoustic signal from multiple sources but did not test by measuring the rotation speed of machines that generate its rotation $\mathscr{D}$-frequency accompanied by several harmonics, that make the extraction of a machine's speed a complicated process.   

\citeauthor{guo2021dancing} \cite{guo2021dancing} propose a mm-wave sensing system that can measure the sub-mm 2D orbit of a rotating machine. The system uses multiple paths of scattered radar signal to regenerate the 2D-orbit of a rotating object. The system assumes prior knowledge of both range of the machine's rotation speed and its vibration band to extract the 2D-trajectory of the rotating machine. These critical assumptions make GWaltz's algorithm unable to measure the machine's rotation speed in the presence of speed harmonics and variable speed conditions. Compared to GWaltz, \AlgoName assumes no prior knowledge about a machine's rotation speed. In all case scenarios, the mean error of \AlgoName is $< 0.5\%$ compared to a calibrated optical tachometer, which proves \AlgoName's high accuracy in machine rotation estimation with no prior knowledge.

\citeauthor{xie2020exploring} \cite{xie2020exploring} propose TagSMM, an RFID system to measure the sub-mm vibration of a machine by measuring corresponding scattered signal phase variation. TagSMM does not need to mount any RFID tags on the machine's body and it requires installing the tag at a specific distance from the RFID reader antenna to satisfy the Fresnel zone theory. The backscattered waves from this tag dampen the static environment interference, enhancing phase variation detection from a vibration source only. TagSMM uses multiple RFID tags to reduce interference from the static environment, increasing deployment costs and making it harder to find ideal locations to install the tags. In an industrial environment, deploying a sensing setup where the RFID reader and tags are located separately and at a specific separation distance is challenging. \AlgoName does not require any reflector's installed on the machines; however, it relies mainly on its signal processing algorithm to adjust its gain and threshold values according to the surrounding environmental conditions. We will design automatic gain and threshold controls in future work, removing any need for manual intervention.

\fakeparagraph{Rotational Doppler Shift.}
Other works that use rotational Doppler shift to measure rotation are \cite{liu2017microwave,gong2018micromotion,zhou2018detection}. In these works, the authors use OAM antennas, horn antennas, rotating smooth discs, and spectrum analysers to measure the speed of a rotating object. However, the experiments were carried out in clean lab implementations, like an anechoic chamber. Such controlled environments can validate the fundamental concept of rotational Doppler shift but ignore the real-world requirements, making the system impractical. Unlike \AlgoName, previous systems did not consider the harmonics created by the rotating blades or gear teeth, which highly affects the rotation speed measurement. Moreover, previous literature's minimum achieved rpm resolution was 60 rpm compared to 1 rpm achieved by \AlgoName. This resolution has been achieved through the combined performance of our proposed two novel signal processing techniques: \AlgoName's speed extraction (Sec.~\ref{speedextraction}) and multiple subcarrier processing algorithms (Sec.~\ref{multiplesubcarriers}). 

In \cite{liu2017microwave,gong2018micromotion}, the authors used a uniform circular array (UCA) of 16 antennas as transmitter and horn antenna as the receiver to generate and receive the OAM waves. The UCA has a diameter of 30 cm, and every antenna needs to receive the signal at a specific phase to generate the OAM wave. To achieve this phase separation, the antennas were connected separately to the signal source through 16 cables making this solution highly impractical. In \cite{zhou2018detection}, the authors used a similar setup to previous works in \cite{liu2017microwave,gong2018micromotion} %
; however, the diameter of the UCA was 80 cm to generate OAM waves with a higher helical twist degree. These setups in generating and receiving OAM waves are not practical in size or complex connection requirements and impose serious difficulty in its practical deployment in industrial environments. The industrial environment has fixed geometry that is not flexible, accommodating sensor setups with a large area, primarily when the Tx and Rx antennas exist in different locations.
Moreover, this work did not validate their solution in real-world scenarios, such as non-line-of-sight or multipath environments.
Compared to previous work, \AlgoName presents a robust and compact monostatic radar sensing system that can measure the rotation speed in the presence of harmonics created by the machine blades and gears. It also has a high resolution of $1$ rpm, which was not possible until now.

\section{Conclusion}
\label{sec:conclusion}

Rotation speed measurement is an essential requirement for the conventional operation of motors, bearings, engines and gears. However, mounting thousands of sensors or reflective tape on those machines is a complex task that increases machine stoppage hours and costs. \AlgoName is a cost-effective, online, contactless and beyond the line-of-sight rotation measurement solution. \AlgoName can measure the rotation speed of several machines simultaneously by using  a combination of OAM waves and lightweight novel signal processing algorithms that give \AlgoName a resolution of 1 rpm, and allows it to operate in high-scattering environments and in presence of dynamic noise. 
Extensive experimental results show the \AlgoName's ability to measure the rotation speed of different machines with metallic and plastic blades, operating at different speeds (up to 7000rpm) and locations (LOS, NLOS and high-scattering environments). Compared to the state-of-the-art tachometers, \AlgoName achieves less than 0.5\% error in most scenarios and a sensing distance of up to 1.2 m with only 150 milliwatts transmit power and a coverage angle of 180 degrees while operating in the 5GHz license-exempt band. 
The low complexity system design of \AlgoName allows it to be implemented in several forms such as a handheld device, wall-mounted or mounted on drones. \AlgoName is a step forward in monitoring rotating machines as it allows for contactless sensing and has demonstrated that it works in several real-world scenarios.

\bibliographystyle{ACM-Reference-Format}
\bibliography{sample-base}
\end{document}